\documentclass[prl,aps,twocolumn,showpacs]{revtex4} 

\usepackage{graphicx}

\begin{document}

\title{Energy Gaps and Roton Structure above the $\nu=1/2$ Laughlin 
State of a Rotating Dilute Bose-Einstein Condensate}

\author{Tatsuya Nakajima$^{1}$ and Masahito Ueda$^{2}$}

\affiliation{
$^1$Physics Department, Graduate School of Science, 
Tohoku University, Sendai 980-8578, Japan \\
$^2$Department of Physics, Tokyo Institute of Technology,
Meguro-ku, Tokyo 152-8551, Japan, \\ and 
CREST, Japan Science and Technology Corporation (JST), 
Saitama 332-0012, Japan}

\date{11 March 2003; published as PRL {\bf 91} (2003) 140401}

\begin{abstract}
Exact diagonalization study of a rotating dilute Bose-Einstein   
condensate reveals that as the first vortex enters the system    
the degeneracy of the low-energy yrast spectrum is lifted and a  
large energy gap emerges. As more vortices enter with faster     
rotation, the energy gap decreases towards zero, but eventually  
the spectrum exhibits a rotonlike structure above the $\nu=1/2$ 
Laughlin state without having a phonon branch despite the        
short-range nature of the interaction.
\end{abstract}

\maketitle

Rotating dilute Bose-Einstein condensates (BECs) have attracted    
considerable interest in recent years~\cite{JILA,MIT,ENS,Oxford}.  
When the single-particle energy-level spacing is much larger than  
the interaction energy per particle, the system becomes highly     
degenerate with increasing the angular momentum (AM) $L$.          
Interactions then play a pivotal role in lifting the degeneracy    
and determining the many-body ground state~\cite{WGS}, in close    
analogy with the physics of the fractional quantum Hall effect     
~\cite{WG,Ho}.
Mottelson~\cite{Mottelson}, on the other hand, pointed out the     
relevance of the yrast line, which traces the lowest-lying states  
of the system as a function of the AM and has served as a key      
concept in nuclear physics~\cite{Hamamoto}, to a rotating dilute   
BEC.
In view of recent development of Feshbach techniques~\cite{IC} and 
experimental achievements of fast rotating BECs~\cite{JILA,MIT,ENS},
such an extreme but highly interdisciplinary arena seems to be     
within experimental reach. 

Bertsch and Papenbrock numerically found that the yrast line       
depends linearly on $L$ for $L \leq N$, where $N$ is the number of 
bosons~\cite{BP99}, and showed that high-lying states above the   
yrast line are dominated by single-particle excitations~\cite{BP01}.
Of particular interest is the yrast spectrum for low-lying states  
because it gives us insights about how the many-body wave function  
responds as vortices enter the system. Mottelson pointed out that  
for $L \ll N$ low-lying states are quasi-degenerate and dominated  
by collective multipolar excitations~\cite{Mottelson}.  An exact   
diagonalization study~\cite{NU00} and subsequent analytical        
studies~\cite{UN01,Kavoulakis,Bardek} have demonstrated the        
existence of the quasi-degenerate yrast spectrum that is dominated 
by interactions between octupole modes.
The purpose of this Letter is to report the results of our exact   
diagonalization study on faster rotating BECs. 
Our primary findings are two distinct kinds of energy gaps: one    
above the single-vortex state and the other associated with        
vortex-antivortex pair excitations above the $\nu=1/2$ Laughlin    
state.
We also find that higher excited states with $L \lesssim N$ 
form linear energy bands. 

Consider a weakly interacting $N$-boson system trapped in an       
axisymmetric parabolic confining potential. 
The potential is assumed to be isotropic in the radial direction   
but in the axial direction it is assumed to be so tightly confined 
that all particles occupy the lowest-energy state in that direction.
The problem is thus essentially two-dimensional with projected AM, 
$L$, on the symmetry axis being conserved.
The many-body Hamiltonian of this system can be written as         
$H = \sum_i h_i + V$, 
where 
$h_i\equiv -\hbar^2 \nabla_i^2/2M+M\omega^2r_i^2/2$ 
denotes the single-particle Hamiltonian for the $i$-th particle and
$V=(4\pi\hbar g/M\omega)\, \sum_{i<j}\delta({\bf r}_i-{\bf r}_j)$  
describes the contact interaction between bosons having mass $M$.  
Here $g$ characterizes the strength of interaction between bosons, 
$g/ \hbar \omega = a_{\rm s}/(2 \pi \hbar/M \omega _z)^{1/2}$,     
$\omega$ and $\omega_z$ are trap frequencies in the radial and     
axial directions, respectively, and $a_{\rm s}$ is the s-wave       
scattering length.
Throughout this Letter we shall assume weak repulsive interactions 
such that $\hbar\omega\gg gN>0$. 

The single-particle state is characterized by the radial quantum   
number $n$ and the AM quantum number $m$, with the corresponding   
eigenenergy given by $\hbar\omega(2n+|m|+1)$. 
For a given AM $L>0$, it is energetically favorable to assume      
$n=0$ and let all particles have nonnegative AM, i.e., $m\geq0$.   
In this lowest-Landau-level approximation, the noninteracting part 
of the Hamiltonian contributes a constant term $\hbar\omega(L+N)$, 
and we shall henceforth ignore this constant part and focus on the 
interaction Hamiltonian.
It is convenient to take as a basis set of states single-particle  
states described by $\phi_m(z)=(z^m/\sqrt{\pi m!})\exp(-|z|^2/ 2)$,
where $z \equiv x + i y$ and the lengths are measured in units of  
$(\hbar/M \omega)^{1/2}$. Expanding the field operator as 
$\hat{\Psi}(z)=\sum_m\hat{b}_m\phi_m(z)$, 
the second-quantized form of the contact interaction Hamiltonian   
is written as 
$\hat{V}=g\sum_{m_1,\cdots,m_4}V_{m_1m_2m_3m_4}\hat{b}_{m_1}^\dagger
\hat{b}_{m_2}^\dagger \hat{b}_{m_3}\hat{b}_{m_4}$,
where the matrix elements $V_{m_1m_2m_3m_4}$ are given by
$V_{m_1 m_2 m_3 m_4} =  \delta _{m_1+m_2, m_3+m_4} \,(m_1+m_2)!/
(2^{m_1+m_2}\,\sqrt{m_1 !\,m_2 !\,m_3 !\,m_4 !}) $. 
Our task thus reduces to finding the energy spectrum of $\hat{V}$ 
under the restrictions of a fixed particle number 
$\sum_m\hat{b}_m^\dagger\hat{b}_m=N$ and a fixed AM 
$\sum_m m \hat{b}_m^\dagger \hat{b}_m=L$.

\begin{figure}[b]
\begin{center}
\includegraphics [width=.9\linewidth]{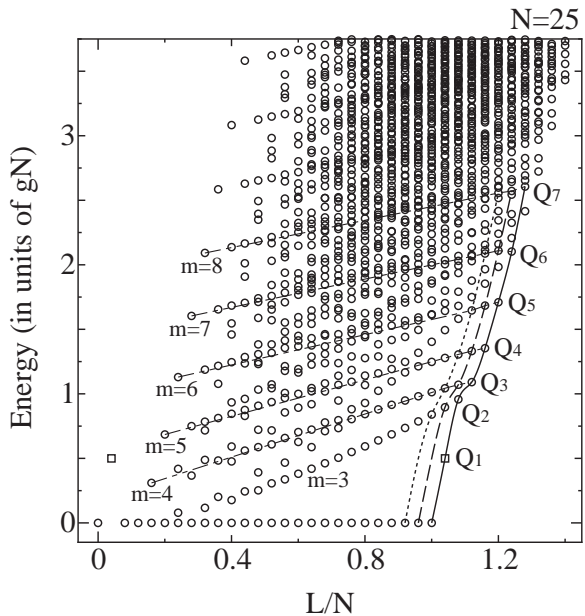}
\end{center}
\vspace*{-0.6cm}
\caption{The yrast spectrum for $0 \leq L \leq 35$ with $N=25$.
As the angular momentum per particle $L/N$ approaches unity, 
excitations associated with single-particle states $\phi_m$  
($m \geq 3$) become increasingly more massive.
The amount of increase in the excitation energy is given in 
Table~\protect\ref{tab1}.
Among rotational bands stabilized by these energy gaps, three 
bands are shown by linking the collective excitations, 
$Q_{\lambda} | L=N  \rangle$ (solid curve),
$Q_{\lambda} | L=N-1\rangle$ (dashed curve), and 
$Q_{\lambda} | L=N-2 \rangle$ (dotted curve), for $\lambda \leq 7$. 
Those states that include excitations of  
the center-of-mass motion are omitted except for the states,  
$Q_1 | L=0 \rangle$ and $Q_1 | L=N \rangle$ (shown by squares). 
The energy is measured from the 
yrast line $E=gN(N-1-L/2)$ in units of $gN$.}
\label{fig1}
\end{figure}

Figure~\ref{fig1} shows the yrast spectrum for $0\leq L\leq 35$
with $N=25$, where the energy is measured in units of $gN$ from
the yrast line $E=gN(N-1-L/2)$~\cite{BP99}. For $L\ll N$, there
are quasi-degenerate excited states arising from pairwise      
repulsive interaction between octupole modes~\cite{NU00}.
However, the excitation energies are very small, of the order  
of $g$.
This quasi-degeneracy may be regarded as a precursor for       
spontaneous symmetry breaking of the axisymmetry associated    
with the entrance of the first vortex. 
Remarkably, the quasi-degeneracy for $L\ll N$ is lifted with   
increasing $L$ and that a large energy gap of the order of     
$gN$ appears as the first vortex enters the system.            
The energy gap implies that the many-body wave function responds
to external rotation in such a manner that a single vortex     
state becomes stabilized.

The emergence of the energy gap can be inferred from the fact  
that excitations associated with the single-particle state     
$\phi_m$ ($m \geq 3$) become increasingly massive, as the AM   
per particle $L/N$ approaches unity.
As shown in Table~\ref{tab1}, the excitation energy increases  
by an amount of the order of $gN$ when the condensate develops 
from the nonvortex state at $L=0$ to the single-vortex state   
at $L=N$, and an increase in excitation energy is larger for   
smaller $m$. In particular, excitations of the $m=3$ mode,     
which are nearly gapless for $L/N \ll 1$, become increasingly  
massive as the vortex enters the system. The energy gap        
continues to increase till $L=N+m-1$ for each $m$.
We also note that the energy bands (shown in Fig.~\ref{fig1}   
as dash-dotted lines) for $m=4,5,6,\cdots$ are almost linear.  
When $L/N \gtrsim 1$, a regular structure of rotational bands  
is stabilized by the energy gaps associated with the 
single-particle states $\phi _m$.
In Figure~\ref{fig1}, we show three rotational bands by linking
the collective excitations $Q_{\lambda}|L=N \rangle$,          
$Q_{\lambda} | L=N-1 \rangle$, and 
$Q_{\lambda} | L=N-2 \rangle$ ($\lambda \leq 7$) 
with solid, dashed, and dotted curves, respectively.   Here    
$Q_\lambda=(1/\sqrt{N\lambda!})\sum_{p=1}^N z_p^\lambda$       
describes the collective multipolar excitation of order        
$\lambda$~\cite{Mottelson}, and 
$|L=N \rangle$, $|L=N-1\rangle$, and $|L=N-2\rangle$ denotes   
the yrast states for $L=N$, $L=N-1$, and $L=N-2$, respectively.

\begin{figure}[t]
\begin{center} 
\includegraphics[width=.85\linewidth]{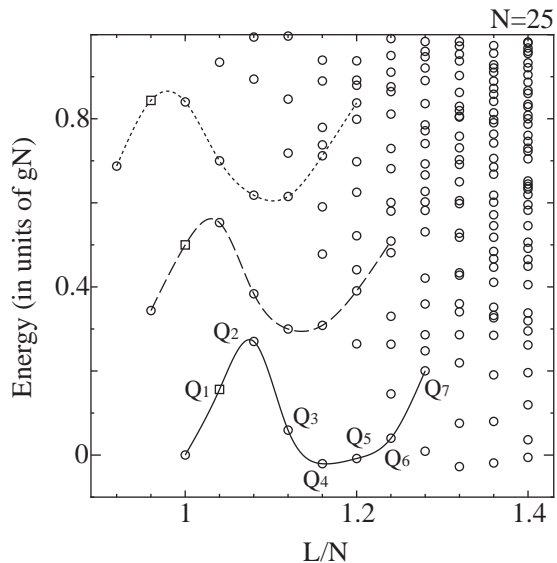}
\end{center} 
\vspace*{-0.6cm}
\caption{Regular rotational bands above the 
single-vortex condensate are shown, where the energy is measured from 
the line $E=E_{L=N}-5gN(L-N)/32$ in units of $gN$.
Among the rotational bands, three bands are shown by linking 
the collective excitations, $Q_{\lambda} | L=N  \rangle$ 
(solid curve), 
$Q_{\lambda} | L=N-1\rangle$ (dashed curve), and 
$Q_{\lambda} | L=N-2\rangle$ (dotted curve), for $\lambda \leq 7$. 
The low-lying excitations can be understood as 
(multiple) excitations of these collective modes. 
Except for the states $Q_1 | L=N \rangle$, $Q_1 | L=N-1 \rangle$,
and $Q_1 | L=N-2 \rangle$ (shown by squares),
those states that include excitations of the center-of-mass 
motion are omitted.}
\label{fig2}   
\end{figure}

The regularity of rotational bands for $L/N \gtrsim 1$ is more 
clearly seen in Fig.~\ref{fig2}, where the energy is measured  
from the line $E=E_{L=N}-5gN(L-N)/32$~\cite{KMP}.
As in Fig.~\ref{fig1}, the collective excitations,             
$Q_{\lambda}|L=N  \rangle$, $Q_{\lambda}|L=N-1\rangle$, and    
$Q_{\lambda}|L=N-2\rangle$ ($\lambda\leq 7$),                  
are linked by solid, dashed, and dotted curves, respectively.  
Those states that contain excitations of the center-of-mass    
motion are not shown except for the states $Q_1|L=N \rangle$,  
$Q_1|L=N-1\rangle$, and $Q_1|L=N-2\rangle$ (shown by squares). 
Other low-lying excitations can be understood as (multiple)    
excitations of these collective modes. 

\begin{figure}[t]
\begin{center} 
\includegraphics[width=.9\linewidth]{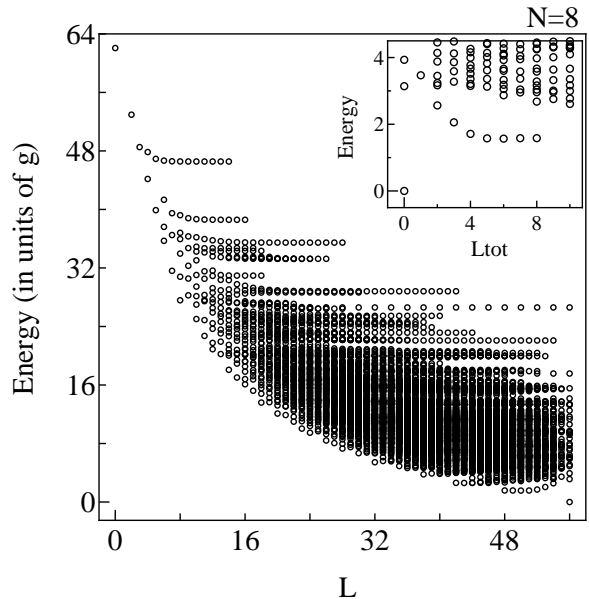}
\end{center} 
\vspace*{-0.6cm}
\caption{The excitation spectrum for $0 \leq L \leq 56$ with $N=8$,
calculated in the spherical geometry.
As the zero-interaction-energy state (i.e., the lowest 
energy state), 
the $\nu = 1/2$ Laughlin state appears at $L=N(N-1)=56$.
In the inset, the excitation spectrum above the 
Laughlin state is shown against the total angular momentum
in the spherical system.
The lowest-energy branch exhibits a rotonlike structure 
without having a phonon branch despite the short-range nature of 
the interaction.}
\label{fig3}   
\end{figure}

When $L\geq N(N-1)$, the system can respond in such a manner   
that the interaction energy becomes zero.                      
In fact, the $\nu=1/2$ Laughlin state~\cite{laughlin},         
$\Psi_{\nu=1/2}=\prod_{p}e^{-|z_p|^2/2}\,\prod_{i<j}(z_i-z_j)^2$,
appears at $L=N(N-1)$ as the first zero-interaction-energy     
state; the contact interaction does not affect this state      
because the minimum of the relative AM between particles is two
for this state. In such a high AM regime, we find a remarkable  
energy gap to appear above the $\nu=1/2$ Laughlin state as     
shown in Fig.~\ref{fig3}.

\begin{table}[b]
\begin{tabular}[t]{|c|c|c|c|c|c|c|}    
\hline 
$\phi _m$  & $\phi _3$ & $\phi _4$ & $\phi _5$ & $\phi _6$ & $\phi _7$ 
  & $\phi _8$ 
\\
\hline  
$\Delta E/gN$  & 1 & 7/8 & 3/4 & 21/32 & 19/32 & 71/128 \\
\hline
\end{tabular}   
\caption{
The increase in energy of the excitation 
associated with the single-particle state $\phi _m$ ($m \geq 3$) 
when the system develops from the nonvortex condensate with $L=0$ 
into the single-vortex condensate with $L=N$.} 
\label{tab1}   
\end{table}    

To investigate long-wavelength collective excitations while  
keeping the average density unaltered (i.e., neutral 
excitations), it is convenient to use the spherical geometry 
~\cite{sphere} rather than the disk geometry which is used   
thus far. The two geometries are related with each other by  
the stereographic mapping.
The $z$-component of the total AM in the spherical geometry  
is given by $L_z = L-NS$, where $2S+1$ is the number of      
single-particle states on the sphere and $2S$ is chosen to be
$2(N-1)$ in order to study the $\nu=1/2$ Laughlin state.     

Figure~\ref{fig3} shows the excitation spectrum calculated   
in the spherical geometry for $0 \leq L \leq 56$ with $N=8$. 
The $\nu = 1/2$ Laughlin state appears at $L=N(N-1)=56$      
(i.e., $L_z =0$) as the zero-interaction-energy state (i.e., 
the lowest energy state) for the contact repulsive           
interaction 
$(4\pi g/R^2)\sum_{i<j}\delta({\bf \Omega}_i-{\bf \Omega}_j)$,
where ${\bf \Omega}_i$ is a unit vector that specifies the   
location of the $i-$th boson on the sphere and the radius of 
the sphere is given by $R=\sqrt{S/2}$.
The total AM, $L_{\rm tot}$, of this state is zero and the   
eigenstates with $L_{\rm tot}=0$ are plotted for $L = NS$.    
The eigenstates with non-zero total AM are plotted only for   
$L=NS-L_{\rm tot}$ without showing their                     
$(2L_{\rm tot}+1)$-fold degeneracy.

The inset of Fig.~\ref{fig3} shows the excitation spectrum   
above the Laughlin state as a function of $L_{\rm tot}$ in   
the spherical system. A rotonlike structure manifests itself
as the lowest-energy excitation branch. 
We have confirmed that the excitation energy remains finite  
($1.229 \pm 0.038$ in units of $g$) in the thermodynamic     
limit~\cite{gaisou}.
We have also confirmed that the feature of the rotonlike    
minimum becomes more pronounced if we use long-range         
interactions such as the Coulomb interaction,                
$\sum_{i<j} |{\bf \Omega}_i - {\bf \Omega}_j|^{-1}$.
The two-roton continuum is also seen to exist above the      
single-roton branch.

To understand the physics underlying these gapful excitations,
let us recall that it is possible to produce quasi-particle   
excitations upon the Laughlin state~\cite{laughlin}.          
One of them is the quasi-hole (vortex) excitation,            
$\prod_{p}[e^{-|z_p|^2/2} \,(z_p-z_0)] \,\prod_{i<j}(z_i-z_j)^2$,
and another one is the `quasi-boson' (anti-vortex) excitation,
$\prod_{p} [e^{-|z_p|^2/2} \,(\frac {\partial \ \,}           
{\partial z_p}-z_0^*)] \,\prod _{i<j} (z_i-z_j)^2$,           
where $z_0$ denotes the location of the vortex or that of the 
anti-vortex.
Then the lowest-lying neutral excitations are considered to be
the bound states of a vortex-antivortex pair.

By investigating a system with $2S=2N-3$ or $2S=2N-1$, we find
that in the thermodynamic limit the quasi-boson creation      
energy remains finite ($1.248 \pm 0.007$ in units of $g$)     
~\cite{gaisou}, while the quasi-hole creation energy is zero  
for repulsive contact interaction.
Since each of quasiboson and quasihole excitations has AM     
$N/2$, the branch composed of a quasiboson-quasihole pair is  
expected to end at $L_{\rm tot}=N$. This is indeed the case in
the lowest-lying excitation branch as shown in Fig.~4, where  
the branches for system size $N=4,5,6,7$ and $8$ are plotted  
against ``wave number'' $k = L_{\rm tot}/R$.

In Fig.~4 it is seen that the minimum of the neutral-excitation
energy (indicated by closed symbols) is realized at            
$k_{\rm min} \simeq 3 \,(M \omega/ \hbar)^{1/2}$.
For $k \geq k_{\rm min}$, the excitation energy increases with
increasing $k$ and approaches the quasi-boson creation energy 
(indicated on the right side of the figure by a line for each 
$N$).
As shown in the inset of Fig.~4, the difference between the   
quasi-boson creation energy (indicated by open circles) and   
the minimum value of the neutral-excitation energy (indicated 
by closed circles) decreases as the system size increases and 
may vanish in the thermodynamic limit. 
In the presence of this finite energy gap, however, this 
rotonlike minimum can survive even for very large values of  
$N$ by introducing, e.g., laser-induced dipole-dipole interactions 
between bosons~\cite{dd}. 

\begin{figure}[t]
\begin{center} 
\includegraphics[width=.9\linewidth]{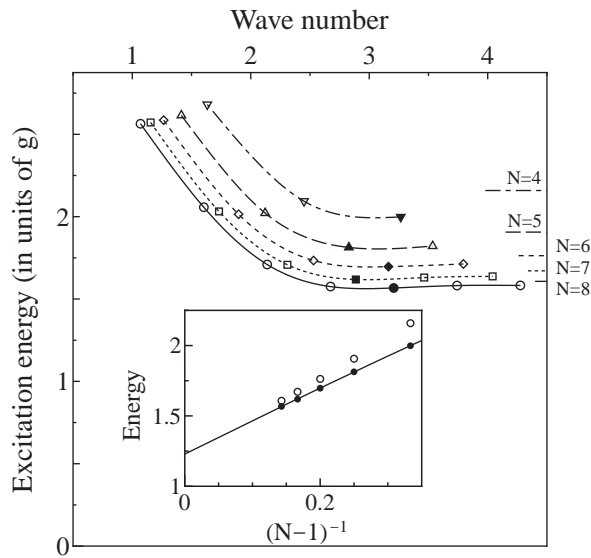}
\end{center} 
\vspace*{-0.6cm}
\caption{The lowest-lying rotonlike excitation branch is shown 
against ``wave number'' $k = L_{\rm tot}/R$ for $N=4,5,6,7$ and $8$. 
The minimum of the neutral-excitation energy is realized 
at $k_{\rm min} \simeq 3 \,(M \omega/ \hbar)^{1/2}$ 
(indicated by closed symbols), and each branch ends at 
$L_{\rm tot}=N$ as expected for a quasihole-quasiboson pair 
excitation. 
For $k \geq k_{\rm min}$, the excitation energy increases 
with increasing $k$ and approaches the quasi-boson creation energy 
(indicated on the right side of the figure by a line for each $N$).
In the inset, the quasiboson creation energy and 
the minimum value of the neutral-excitation energy are shown 
against $1/(N-1)$ 
by open and closed circles, respectively.}
\label{fig4}   
\end{figure}   

It should be noted that the lowest-lying spectrum has no     
phonon branch, despite the short-range nature of the         
interaction~\cite{rj}.
A similar roton spectrum has been reported in the fractional 
quantum Hall system~\cite{GMP}, where absence of the phonon  
branch has been attributed to the Higgs mechanism caused by  
the long-range nature of the Coulomb interaction.
However, the rotonlike structure with no phonon branch found 
in this Letter for the contact interaction does not fit into 
this category. 
In the composite-boson picture, the $\nu=1/2$ Laughlin state 
is nothing but a condensate of composite bosons, each of     
which is composed of a boson and two flux quanta attached to 
it.
In the composite-particle picture for the quantum Hall       
system, it has been discussed that interactions between      
composite particles are different from those between original 
particles~\cite{devna}. It might be conceived that the finite 
gap in the long-wavelength limit is related with             
momentum-dependent long-range interactions introduced on the 
composite-boson transformation (without being induced  
by a laser~\cite{dd}).
However, the understanding of the gap formation found in this 
Letter remains to be clarified and merits further investigation.

T.N. and M.U. acknowledge support by Grant-in-Aids for       
Scientific Research (Grant No.14740181 and No.15340129) by   
the Ministry of Education, Culture, Sports, Science and 
Technology of Japan, respectively.
M.U. also acknowledges support by the Toray Science Foundation 
and by the Yamada Science Foundation.


\begin{thebibliography}{99}

\bibitem{JILA} 
P.C. Haljan {\it et al}, 
Phys. Rev. Lett. {\bf 87}, 210403 (2001); 
P. Engels {\it et al}, 
{\it ibid.} {\bf 89}, 100403 (2002).

\bibitem{MIT}
J.R. Abo-Shaeer {\it et al}, 
Science {\bf 292}, 476 (2001).

\bibitem{ENS}
P. Rosenbusch {\it et al}, 
Phys. Rev. Lett. {\bf 88}, 250403 (2002).

\bibitem{Oxford}
E. Hodby {\it et al}, 
Phys. Rev. Lett. {\bf 88}, 010405 (2002).

\bibitem{WGS} N.K. Wilkin {\it et al}, 
Phys. Rev. Lett. {\bf 80}, 2265 (1998).

\bibitem{WG} N.K. Wilkin and J.M.F. Gunn, 
Phys. Rev. Lett. {\bf 84}, 6 (2000).

\bibitem{Ho} T.-L. Ho, Phys. Rev. Lett. {\bf 87}, 060403 (2001).

\bibitem{Mottelson} B. Mottelson, 
Phys. Rev. Lett. {\bf 83}, 2695 (1999).

\bibitem{Hamamoto} I. Hamamoto, 
in {\it Treatise on Heavy-Ion Science},
edited by D.A. Bromley (Plenum, New York, 1985), vol.3, p.313;
I. Hamamoto and B. Mottelson, Nucl. Phys. {\bf A507}, 65 (1990).

\bibitem{IC} 
S. Inouye {\it et al}, 
Nature {\bf 392}, 151 (1998);
S.L. Cornish {\it et al}, 
Phys. Rev. Lett. {\bf 85}, 1795 (2000).

\bibitem{BP99} G.F. Bertsch and T. Papenbrock,
Phys. Rev. Lett. {\bf 83}, 5412 (1999).

\bibitem{BP01} T. Papenbrock and G.F. Bertsch, 
Phys. Rev. A {\bf 63}, 023616 (2001).

\bibitem{NU00} T. Nakajima and M. Ueda, 
Phys. Rev. A {\bf 63}, 043610 (2001).

\bibitem{UN01} M. Ueda and T. Nakajima, 
Phys. Rev. A {\bf 64}, 063609 (2001).

\bibitem{Kavoulakis} G.M. Kavoulakis {\it et al}, 
Phys. Rev. A {\bf 63}, 055602 (2001).

\bibitem{Bardek} V. Bardek {\it et al}, 
Phys. Rev. A {\bf 64}, 015603 (2001); 
V. Bardek and S. Meljanac, {\it ibid.} 
{\bf 65}, 013602 (2002).

\bibitem{KMP} G.M. Kavoulakis {\it et al}, 
Phys. Rev. A {\bf 62}, 063605 (2000).

\bibitem{laughlin} R.B. Laughlin, 
Phys. Rev. Lett. {\bf 50}, 1395 (1983).

\bibitem{sphere} F.D.M. Haldane, 
Phys. Rev. Lett. {\bf 51}, 605 (1983); 
G. Fano {\it et al}, 
Phys. Rev. B {\bf 34}, 2670 (1986).

\bibitem{gaisou} 
The best fit result for the neutral excitation gap is 
$1.229 +2.394/(N-1)-0.2476/(N-1)^2$ in units of $g$, 
while the one for the quasi-boson creation energy is
$1.248 +2.367/(N-3/2)-0.2279/(N-3/2)^2$.

\bibitem{dd} D.H.J. O'Dell {\it et al}, 
Phys. Rev. Lett. {\bf 90}, 110402 (2003); 
L. Santos {\it et al}, 
{\it ibid.} {\bf 90}, 250403 (2003).

\bibitem{rj} In an unpublished work by N. Regnault 
and Th. Jolicoeur (cond-mat/0212477), a related subject on the 
finiteness of neutral-excitation energy gap is discussed.

\bibitem{GMP} S.M. Girvin {\it et al}, 
Phys. Rev. Lett. {\bf 54}, 581 (1985).

\bibitem{devna} G. Dev and J.K. Jain, 
Phys. Rev. Lett. {\bf 69}, 2843 (1992); 
T. Nakajima and H. Aoki, {\it ibid.} {\bf 73}, 3568 (1994).
On the theory of composite particle as a dipole, 
see R. Shankar and G. Murthy, 
{\it ibid.} {\bf 79}, 4437 (1997) and references therein.

\end{thebibliography}
\end{document}